\documentclass[12pt,a4paper]{article}
\usepackage{times}
\usepackage{amssymb}
\usepackage{latexsym}
\usepackage{amsmath}

\numberwithin{equation}{section}

\newcommand{\abstracttitle}[1]{\parbox{\textwidth}
                              {\setlength{\baselineskip}{17pt}%
                               \begin{center}{\bfseries #1}\end{center}}}
\newcommand{\authors}[1]{\begin{center}#1\end{center}}
\newcommand{\addresses}[1]{\begin{center}#1\end{center}}

\begin{document}
\abstracttitle
{HYPERFINE SPLITTING IN HEAVY IONS WITH THE NUCLEAR MAGNETIZATION
DISTRIBUTION DETERMINED FROM EXPERIMENTS ON MUONIC ATOMS}

\authors{A.~A.~Elizarov, 
V.~M.~Shabaev \footnote{ Corresponding author. Fax: +007 812 4287240.
E-mail: shabaev@pcqnt1.phys.spbu.ru}, 
N.~S.~Oreshkina, and I.~I.~Tupitsyn}

\addresses{
Department of Physics, St.Petersburg State University, Oulianovskaya 1,
Petrodvorets, St.Petersburg 198504, Russia
\\
}

\abstract
{
The hyperfine
splitting in hydrogenlike $^{209}$Bi, $^{203}$Tl, and $^{205}$Tl 
is calculated with the nuclear magnetization  determined from
 experimental data on the hyperfine splitting in the corresponding muonic
atoms. The single-particle and configuration-mixing nuclear models 
are considered. The QED corrections are taken into account for both
electronic and muonic atoms.  
The obtained results are compared
with other calculations and with experiment.  
\newline
\,\,
\newline
{\small PACS numbers: 12.20.Ds, 31.30.Jv, 31.30.Gs} \newline
{\small Keywords: hyperfine splitting, heavy ions, quantum electrodynamics}
}

\setlength{\parskip}{\bigskipamount}

\section{Introduction}\label{S:intro}

High-precision measurements of the hyperfine splitting (HFS) 
in heavy hydrogenlike ions~\cite{exp1,exp2,exp3,exp4,exp5}
have triggered a great interest to calculations of this effect.
The main goal of these experiments was to probe the magnetic sector of
quantum electrodynamics (QED) in the presence of a strong Coulomb field.
The uncertainty of the theoretical results is mainly determined
 by the uncertainty of the nuclear magnetization distribution correction,
 the so-called Bohr-Weisskopf (BW) effect. In calculations, based on 
the single-particle nuclear model~\cite{PhB,Sh97,Spr,Lab}, which provide a reasonable agreement
with the experiments, this
 uncertainty may amount up to about 100\% of the BW effect
and is generally few times larger than the total QED contribution.
More elaborated calculations, based on many-particle nuclear models~\cite{Tom,DmSe}, do not provide
a desirable agreement with the experiments.

In the present paper, we determine the BW correction to the hyperfine
splitting of hydrogenlike $^{209}$Bi, $^{203}$Tl, and $^{205}$Tl 
using experimental data on the hyperfine splitting in the corresponding muonic
atoms. We consider the single-particle and configuration-mixing nuclear models.  
The parameters of the nuclear magnetization distribution are chosen
to reproduce the experimental values of the nuclear magnetic moment
as well as the BW contribution in  muonic atoms extracted from the corresponding
experiments. To increase the precision of determining the BW contribution,  
the QED corrections for the muonic atoms have been evaluated. 
The obtained results are compared
with other calculations and with experiment.  

The relativistic units ($\hbar=c=1$) and the Heaviside charge unit 
($\alpha=e^2/(4\pi)$, $e<0$) are used in the paper.

\section{Hyperfine splitting in muonic atoms}\label{s:ked}
The ground-state hyperfine splitting in muonic atoms can be written in the form:
\begin{equation}\label{E:STS}
\varDelta E=\varDelta E_{\mathrm{NS}}+\varDelta E_{\mathrm{BW}}+\varDelta E_{\mathrm{QED}},
\end{equation}
where $\varDelta E_{\mathrm{NS}}$ is the hyperfine splitting value 
incorporating the relativistic and nuclear charge distribution ("nuclear size") effects, 
$\varDelta E_{\mathrm{BW}}$ is the BW 
contribution, $\varDelta E_{\mathrm{QED}}$ is the QED correction. 
The $\varDelta E_{\mathrm{NS}}$ value can be calculated by the formula:
\begin{equation}\label{E:NS}
\varDelta E_{\mathrm{NS}}=-\alpha\frac{4}{3}\frac{\mu}{\mu_{N}}\frac{1}{m_p}\frac{(2I+1)}{2I}
\int\limits_{0}^{\infty}dr \: g(r)f(r),
\end{equation}
where $\alpha$ is the fine structure constant, $\mu$ is the nuclear magnetic 
moment, $\mu_N$ is the nuclear magneton, $m_p$ is the proton mass, and $I$ is the nuclear spin. $g(r)$ and $f(r)$ 
are the radial parts of the Dirac wave function:
\begin{equation}
\Psi(\textbf{r})=
\begin{pmatrix}
	g(r)\varOmega_{\kappa m}(\textbf{n}) \\
	\mathit{i}f(r)\varOmega_{-\kappa m}(\textbf{n})
\end{pmatrix},
\end{equation}
which are determined  by solving 
the Dirac equation 
with the Fermi distribution of the nuclear charge ($4\pi\int dr r^{2}\rho(r)=1$):
\begin{equation}
\rho (r)=\frac{\rho_{0}}{1+\mathrm{exp}(\frac{r-c}{a})}.
\end{equation}
Here $c$ is the half-density radius and $a$ is related to the skin thickness $t$ by $t=(4\log 3)a$, 
defined as the distance over which the charge density falls from 90\% to 10\% of its maximum value. 

The individual contributions to $\varDelta E$ for muonic atoms of $^{203}$Tl, 
$^{205}$Tl, and $^{209}$Bi are presented in Table \ref{ta:mu}. 
The $\varDelta E_{\mathrm{NS}}$ values are given in the second column. 
In the third column we present the BW correction evaluated within the single-particle nuclear model 
according to the prescriptions given in \cite{PhB,Sh97,Spr} 
(see also the next section of the present paper). The wave function of the odd nucleon was determined by 
solving the Schr\"odinger equation with the Woods-Saxon potential. 
In muonic atoms, the QED correction is mainly determined by the vacuum polarization (VP) contribution,
which consists of the electronic electric-loop and magnetic-loop parts. 
In the Uehling approximation, the electric-loop part is determined by the potential:
\begin{align}\label{E:U1}
U_{\mathrm{Ue}}^{\mathrm{EL}}(r)=-\alpha Z\frac{2\alpha}{3\pi}\int\limits_{0}^{\infty}dr'
4\pi \rho (r')\int\limits_{1}^{\infty}dt\left(1+\frac{1}{2t^{2}}\right)
\frac{\sqrt{t^{2}-1}}{t^{2}}\\ \notag
\times\frac{\mathrm{exp}(-2m|r-r'|t)-\mathrm{exp}(-2m(r+r')t)}{4mrt} ,
\end{align}
where $m$ is the electron mass.
The corresponding correction ($\varDelta E_{\mathrm{VP}}^{\mathrm{EL}}$) is 
derived as the difference of equations (\ref{E:NS}) 
with the wave functions obtained by solving the Dirac equation with and without the Uehling potential (\ref{E:U1}).
For the correction to the hyperfine splitting due to the magnetic loop one obtains:
\begin{equation}
\varDelta E_{\mathrm{VP}}^{\mathrm{ML}}=\langle A|U_{\mathrm{VP}}^{\mathrm{ML}}(\textbf{r})|A\rangle,
\end{equation}
where $|A\rangle$ is the state vector of the whole atomic system,
\begin{align}
U_{\mathrm{VP}}^{\mathrm{ML}}(\textbf{r})=H_{\mathrm{hfs}}(\textbf{r})
\frac{2\alpha}{3\pi}\int\limits_{1}^{\infty}
dt\left(1+\frac{1}{2t^{2}}\right)
\frac{\sqrt{t^{2}-1}}{t^{2}} 
\medspace(1+2mrt)\medspace\mathrm{exp}(-2mrt),
\end{align}
$H_{\mathrm{hfs}}(\textbf{r})$ is the hyperfine interaction operator:
\begin{align}\label{E:Hhfs}
H_{\mathrm{hfs}}(\textbf{r})=\frac{|e|}{4\pi}\frac{(\boldsymbol{\alpha}\cdot[\boldsymbol{\mu}
\times\mathbf{r}])}{r^{3}},
\end{align}
$\boldsymbol{\mu}$ is the nuclear-magnetic-moment operator, and $\boldsymbol{\alpha}$ is a vector 
incorporating the Dirac matrices.
The total QED correction ($\varDelta E_{\mathrm{QED}}\approx 
\varDelta E^{\mathrm{EL}}_{\mathrm{VP}}+\varDelta E^{\mathrm{ML}}_{\mathrm{VP}}$) 
is given in the fourth column of Table \ref{ta:mu}. The total theoretical values obtained in this work 
(fifth column) are in a good agreement with the experimental ones 
(sixth column) and with the previous theoretical calculations \cite{JS}, which do not 
account for the QED corrections. Since the theoretical uncertainty is 
mainly determined by the uncertainty of the BW effect, the experimental values of the hyperfine 
splitting in muonic atoms can be employed to determine the BW contribution and, 
therefore, the parameters of the nuclear magnetization distribution for a given 
nuclear model. Then, with these parameters, the BW correction to the hyperfine splitting 
in the corresponding electronic ions can be calculated. Such 
calculations are presented in the third and fourth sections for the single-particle 
and configuration-mixing nuclear models, respectively.
\begin{table}
	\caption{Individual contributions to the hyperfine splitting in muonic atoms, in keV.}
\label{ta:mu}
\begin{center}
	\begin{tabular}{lccccc}
	\hline\hline  Atom &$\varDelta E_{\mathrm{NS}}$ & $\varDelta E_{\mathrm{BW}}$
	&$\varDelta E_{\mathrm{QED}}$& 
	$\varDelta E_{\mathrm{theor}}$&
	$\varDelta E_{\mathrm{exp}}^{\mathrm{\mu^{-}}}$
	 \\
	\hline
        $^{203}$Tl & 4.70 & -2.06(66) & 0.06 & 2.70(66) & 2.66(30)~\cite{expmu1}  \\
        $^{205}$Tl & 4.73 & -2.06(66) & 0.06 & 2.73(66) & 2.32(6)~\cite{expmu1}  \\
	$^{209}$Bi & 6.69 & -2.49(80) & 0.09 & 4.29(80) & 4.44(15)~\cite{expmu2} \\
	\hline
	\end{tabular}	
\end{center}
\end{table}
\section{Nuclear magnetization in the single-particle model}\label{s:sp}
In the single particle model, the nuclear magnetization is ascribed to the odd nucleon.
The nuclear magnetic moment is given by
\begin{align}\label{A:mumu1a}
\frac{\mu}{\mu_N}=
\begin{cases}
	\frac{1}{2}\left[g_S+(2I-1)g_L\right], &\text{$I=L+1/2$}\\
	\frac{I}{2(I+1)}\left[-g_S+(2I+3)g_L\right], &\text{$I=L-1/2$},
\end{cases}
\end{align}
where $I$ and $L$ are the total and orbital angular momenta of the odd nucleon. For proton $g_L=1$ and
for neutron $g_L=0$. 
In calculations of the HFS the $g_S$ factor is usually chosen to yield 
the observed value of the nuclear magnetic moment.
To calculate the BW effect within the single-particle model, one has to adopt the replacement 
$\mu \rightarrow \mu(r)=F(r)\mu$ in the HFS operator.
The function $F(r)$ is given by~\cite{Spr}:
\begin{eqnarray}
F(r)&=&\frac{\mu_N}{\mu}\Bigl\{
\Bigl[\frac{1}{2}g_S+\Bigl(I-\frac{1}{2}\Bigr)g_L\Bigr]
\int\limits_0^r dr'r'^2 u^2(r')\nonumber\\
&& +\left[-\frac{2I-1}{8(I+1)}g_S+\left(I-\frac{1}{2}\right)g_L\right]
\int\limits_r^{\infty}dr'r'^2 u^2(r') \left(\frac{r}{r'}\right)^3 \Bigr\}
\end{eqnarray}
for $I=L+1/2$ and 
\begin{eqnarray}
F(r)&=&\frac{\mu_N}{\mu}\Bigl\{
\Bigl[-\frac{I}{2(I+1)}g_S+\frac{I(2I+3)}{2(I+1)}g_L\Bigr]
\int\limits_0^r dr'r'^2 u^2(r')\nonumber\\
&&+\left[\frac{2I+3}{8(I+1)}g_S+\frac{I(2I+3)}{2(I+1)}g_L\right]
\int\limits_r^{\infty}dr'r'^2 u^2(r') \left(\frac{r}{r'}\right)^3\Bigr\} 
\end{eqnarray}
for $I=L-1/2$. 
Here $u(r)$ is the radial part of the wave function of the odd nucleon. 
The relative value of the BW correction, defined by 
$\varepsilon=-\varDelta E_{\mathrm{BW}}/\varDelta E_{\mathrm{NS}}$, 
can be written in the form:
\begin{equation}
\varepsilon_{\mathrm{s.p.}}=1-\frac{\int\limits_{0}^{\infty}drF(r)g(r)f(r)}{\int\limits_{0}^{\infty}drg(r)f(r)}.
\end{equation}
To reproduce the experimental values for the HFS in muonic atoms within the single-particle nuclear model, 
we have considered the following parameterization for $u(r)$:
\begin{align}\label{A:volnn2}
\begin{cases}
	u(r)=u_0r^n, \: n=0,1,2, &\text{$ r \leqslant R_{\rm M} $}\\
	u(r)=0, &\text{$r > R_{\rm M}$}
\end{cases}
\end{align}
and
\begin{align}\label{A:volnn3}
\begin{cases}
	u(r)=u_0(R_{\rm M}-r)^n,\: n=1,2 , &\text{$ r \leqslant R_{\rm M} $}\\
	u(r)=0, &\text{$r > R_{\rm M}$.}
\end{cases}
\end{align} 
The constant $u_0$ is determined by: 
$\int_0^{\infty}drr^2u^2(r)=1$.
The magnetic radius $R_{\rm M}$ is derived from the equation:
\begin{equation}\label{E:fR}
\varDelta E_{\mathrm{exp}}^{\mu^{-}}-
\varDelta E_{\mathrm{QED}}=(1-\varepsilon_{\mathrm{s.p.}}(R_{\rm M}))\varDelta E_{\mathrm{NS}},
\end{equation}
where $\varDelta E_{\mathrm{exp}}^{\mathrm{\mu^{-}}}$ and $\varDelta E_{\mathrm{QED}}$ 
are the experimental value of the HFS for muonic atom and the QED correction, accordingly. 
Then we can calculate the BW correction for 
the corresponding electronic ions with $R_{\rm M}$ derived from equation (\ref{E:fR}).
We have found that the results for the BW correction ($\varepsilon_{\mathrm{s.p.}}$) 
in electronic H-like ions are stable enough within 
the parameterizations (\ref{A:volnn2}), (\ref{A:volnn3}). For this reason, the uncertainties 
of $\varepsilon_{\mathrm{s.p.}}$ presented in 
Table \ref{ta:xxx}, are mainly determined by the uncertainty of the experimental HFS values in muonic atoms.
\begin{table}
	\caption{The Bohr-Weisskopf correction to the HFS in electronic hydrogenlike ions, 
derived from experiments on 
	muonic atoms within the single-particle nuclear model.
	$\delta\varepsilon_{\mathrm{s.p.}}^{\mathrm{mod.}}$ 
denotes the uncertainty of $\varepsilon_{\mathrm{s.p.}}$ 
	caused by the different parameterizations used in the calculation 
	(see equations (\ref{A:volnn2}), (\ref{A:volnn3})).}\label{ta:xxx}
	\begin{center}
	\begin{tabular}{lcccc}
\hline\hline           
	Ion&    $I^\pi$   &$\varDelta E_{\mathrm{exp}}^{\mathrm{\mu^{-}}}$(keV)&$\varepsilon_{\mathrm{s.p.}}$
	&$\delta\varepsilon_{\mathrm{s.p.}}^{\mathrm{mod.}}$ \\
\hline  
       $^{203}$Tl$^{80+}$ &$\frac{1}{2}+$& 2.66(30)~\cite{expmu1}& 0.0155(35)&9\% \\
       $^{205}$Tl$^{80+}$ &$\frac{1}{2}+$& 2.32(6) ~\cite{expmu1}& 0.0193(24)&11\% \\
       $^{209}$Bi$^{82+}$ &$\frac{9}{2}-$& 4.44(15)~\cite{expmu2}& 0.0123(14)&8\%  \\
\hline\hline
	
	\end{tabular}
	\end{center}	
\end{table}
\section{Nuclear magnetization in the configuration-mixing model}\label{s:mc}
In the configuration-mixing model the nuclear magnetism is determined by the last odd nucleon and the particle-hole 
excited states. 
So, the HFS (without the QED correction) can be represented as:
\begin{equation}\label{e:msm}
	\varDelta E_{\mathrm{c.m.}}=\varDelta E_{\mathrm{s.p.}}+\delta E(\varDelta\mu),
\end{equation}
where $\varDelta\mu$ is the correction to the nuclear magnetic moment, due to mixing particle-hole states.
The nuclear magnetic moment can be written in the form:
\begin{equation}
\mu_{\mathrm{exp}}=\mu_{\mathrm{s.p.}}+\varDelta\mu.
\end{equation}
The formulas for $\varDelta E_{\mathrm{s.p.}}$ are well known (see, e.g.,~\cite{JS}):
\begin{align}
\varDelta E_{\mathrm{s.p.}}=
-\alpha\frac{4}{3}\frac{1}{m_p}\frac{2I+1}{2I} \Bigl \{ \Bigl[\frac{1}{2}g_S+\Bigl(I-\frac{1}{2}\Bigr)g_L\Bigr]
K_a+ \nonumber\\ 
+\Bigl[-\frac{2I-1}{8(I+1)}g_S+\Bigl(I-\frac{1}{2}\Bigr)g_L\Bigr]K_b\Bigr\}, &\text{ \:$I=L+1/2$},
\end{align}
\begin{align}
\varDelta E_{\mathrm{s.p.}}=-\alpha\frac{4}{3}\frac{1}{m_p}\frac{2I+1}{2I}
\Bigl\{
\Bigl[-\frac{I}{2(I+1)}g_S+\frac{I(2I+3)}{2(I+1)}g_L\Bigr]
K_a+ \nonumber \\ 
+\Bigl[\frac{2I+3}{8(I+1)}g_S+\frac{I(2I+3)}{2(I+1)}g_L\Bigr]K_b\Bigr\}, &\text{ \:$I=L-1/2$},
\end{align}
where
\begin{equation}
	K_{a} = \int\limits_{0}^{\infty} dR\: R^2 u_{n,L}^{2} (R)
		\int\limits_{R}^{\infty} dr\: g(r)f(r),
\end{equation}
\begin{equation}
	K_{b} = \int\limits_{0}^{\infty} dR\: R^{-1}u_{n,L}^{2}(R)
		\int\limits_{0}^{R} dr\: g(r)f(r).
\end{equation}
The correction for the $\Delta L=0$ mixing terms is given by~\cite{JS}:
\begin{eqnarray}
\delta E(\varDelta\mu)&=&-\alpha\frac{4}{3}\frac{1}{m_p}\frac{2I+1}{2I}\int\limits_{0}^{\infty} dr \: f(r)g(r)
\sum_{L^{'}}\zeta_{L^{'}}\varDelta\mu\nonumber\\
&&\times\left[\{1-K_S\}_{L^{'}}+ 
 \frac{\frac{1}{4}g_S-g_L}{g_S-g_L}
		\{ K_{S}-K_{L} \}_{L^{'}}\right],
\end{eqnarray}
\begin{equation}
	\sum_{L^{'}}\zeta_{L^{'}}=1.
\end{equation}
The quantities $\zeta_{L^{'}}$ determine the re-distribution $\varDelta\mu$ between the configurations.
We use the following designations:
\begin{equation}
	\{A\}_{L^{'}} = \int\limits_{0}^{\infty}dR\: R^2 u_{n,L^{'},J}(R) u_{n,L^{'},J^{'}}(R) 
		A(R), 
\end{equation}
\begin{equation}
	K_S(R)=\frac{\int\limits^R_0 dr \: g(r)f(r)}{\int\limits^{\infty}_0 dr \: g(r)f(r)},
\end{equation}
\begin{equation}
	K_L(R)=\frac{\int\limits^R_0 dr\:(1-\frac{r^3}{R^3})g(r)f(r)}{\int\limits^{\infty}_0 dr \:g(r)f(r)}.
\end{equation}
Here $u_{n,L}$ and $u_{n,L,J}$ are the radial parts of the wave functions of the odd nucleon. 
They are obtained by solving the Schr\"odinger equation with the Woods-Saxon potential.
We consider $\zeta_{L^{'}}=\frac{1}{M}$, where $M$ is the number of the configurations.
The quantities $\varDelta\mu$ and $g_S$ are derived from the equations:
\begin{align}\label{a:mg}
	\frac{\mu_{\mathrm{exp}}}{\mu_N}=
\begin{cases}
\frac{1}{2}g_S+\left(I-\frac{1}{2}\right)g_L+
\frac{\varDelta\mu}{\mu_N}\sum_{L^{'}}\zeta_{L^{'}}\{1\}_{L^{'}}
, &\text{$I=L+1/2$}\\
	-\frac{I}{2\left(I+1\right)}
g_S+\frac{I\left(2I+3\right)}{2\left(I+1\right)}g_L+
\frac{\varDelta\mu}{\mu_N}\sum_{L^{'}}\zeta_{L^{'}}\{1\}_{L^{'}}
, &\text{$I=L-1/2$},
\end{cases}
\end{align}
\begin{equation}\label{e:e}
\varDelta E_{\mathrm{exp}}^{\mathrm{\mu^{-}}}-\varDelta E_{\mathrm{QED}}=\varDelta E_{\mathrm{s.p.}}
(g_S)+\delta E\left(\varDelta\mu\right).
\end{equation}
The Bohr-Weisskopf correction is determined by:
\begin{equation}
1-\varepsilon_{\mathrm{c.m.}}=
\frac{\varDelta E_{\mathrm{s.p.}}(g_S)+\delta E(\varDelta\mu)}{\varDelta E_ {\mathrm{NS}} }.
\end{equation}
With $g_S$ and $\varDelta\mu$ derived from equations (\ref{a:mg}), (\ref{e:e}), we can calculate the BW correction 
for the electronic ions.
The results of these calculations are presented in Table 
\ref{ta:bwmc}. The values of $\varepsilon_{\mathrm{c.m.}}$ given in 
Table \ref{ta:bwmc} are in a good agreement with $\varepsilon_{\mathrm{s.p.}}$ presented in Table \ref{ta:xxx}.
\begin{table}
	\caption{The Bohr-Weisskopf correction to the HFS in electronic hydrogenlike ions, derived from experiments 
on muonic atoms within the configuration-mixing model. 
	}
	\label{ta:bwmc}
	\begin{center}
	\begin{tabular}{lllllll}
\hline\hline           
	Ion&    $I^\pi$   &
      $\varDelta E_{\mathrm{exp}}^{\mathrm{\mu^{-}}}$(keV)&$\varepsilon_{\mathrm{c.m.}}$
\\
\hline 
	$^{203}$Tl$^{80+}$ &$\frac{1}{2}+$& 2.66(30)~\cite{expmu1}& 0.0179(36) & \\
	$^{205}$Tl$^{80+}$ &$\frac{1}{2}+$& 2.32(6)~\cite{expmu1}&  0.0214(6) & \\
        $^{209}$Bi$^{82+}$ &$\frac{9}{2}-$& 4.44(15)~\cite{expmu2}& 0.0119(11)  & \\
\hline\hline
	
	\end{tabular}
	\end{center}	
\end{table}

\section{Results and discussion}
In Table \ref{ta:bw} we compare the BW correction to the HFS in electronic H-like ions, derived from the 
experiments on muonic atoms. $\varepsilon _{\mathrm{s.p.}}$ and $\varepsilon _{\mathrm{c.m.}}$ are obtained 
employing the single-particle and configuration-mixing models, respectively. $\varepsilon$ is 
a value, which corresponds to the parameterization (\ref{A:volnn2}) with $n=0$ and whose uncertainty 
covers all the $\varepsilon _{\mathrm{s.p.}}$ and $\varepsilon _{\mathrm{c.m.}}$ values. For comparison, 
the values of the BW correction obtained previously by direct calculations within the single-particle and 
many-particle models are presented as well. We conclude that our results for $\varepsilon$ are stable enough with 
respect to a change of the nuclear model. They also have a better accuracy than the previous single-particle 
results~\cite{PhB,Sh97,Spr,Lab}.

In Table \ref{ta:ok} we present our final 
theoretical results for the HFS in electronic hydrogen-like ions. These results, 
which include the BW correction derived in 
the present work and the QED correction taken from \cite{Sh97,VP,VP1,Shabz}, are 
compared with previous calculations and with experiment. 
As one can see from the table, our results are closer to the experimental ones, 
compared to the results based on the direct calculations within the single-particle nuclear model \cite{Spr}.
However, due to a higher accuracy of the present results, a small 
discrepancy between the theory and experiment occurs for 
$^{209}$Bi$^{82+}$. The reason for this discrepancy is unclear to us.

\section*{Acknowledgments}

Valuable conversations with A.N. Artemyev, L.N. Labzowsky, and L. Simons
 are gratefully acknowledged. 
This work was supported in part by RFBR (Grant No.
04-02-17574), by the Russian Ministry of Education (Grant No.
E02-3.1-49),
 and by INTAS-GSI (Grant No. 03-54-3604). 
 N.S.O. acknowledges the support from the Federal Education Agency (grant No. A04-2.9-151).

\begin{table}
	\caption{The Bohr-Weisskopf correction 
to the HFS in electronic H-like ions, derived from experiments on muonic atoms. 
	}\label{ta:bw}
	\begin{center}
	\begin{tabular}{lcllll}
	\hline\hline 
	            &     &          &$^{209}$Bi$^{82+}$&$^{205}$Tl$^{80+}$&$^{203}$Tl$^{80+}$\\
	\hline
	&&$\varepsilon_{\mathrm{s.p.}}$       &0.0123(14)        &0.0193(24)        &0.0155(35) \\
	&&$\varepsilon_{\mathrm{c.m.}}$       &0.0119(11)        &0.0214(6)         &0.0179(36) \\
	&&$\varepsilon                $       &0.0123(15)        &0.0193(27)        &0.0155(40) \\
	\hline
	Shabaev et al. &~\cite{Spr} &$\varepsilon$  &0.0118            &0.0179            &0.0179     \\
	Labzowsky et al.& ~\cite{Lab}&$\varepsilon$   &0.0131            &                  &           \\
	Tomaselli et al. & ~\cite{Tom}&$\varepsilon$  &0.0210            &                  &           \\
	Sen'kov and Dmitriev &~\cite{DmSe}&$\varepsilon$  &0.0095$\binom{+7}{-38}$&&                \\           
	\hline \hline
	\end{tabular}
	\end{center}
\end{table}
\begin{table}
\caption{The total theoretical results for the hyperfine splitting in electronic H-like ions, in eV.}\label{ta:ok}
\begin{center}
\begin{tabular}{cccc}
	\hline\hline
	                     &  Theory     &Theory     & Experiment \\
	                     &[this work]&\cite{Spr} &             \\
	\hline
	$^{203}$Tl$^{80+}$   & 3.220(20)   & 3.229(17) & 3.21351(25) \cite{exp5}                     \\
	$^{205}$Tl$^{80+}$   & 3.238(9)    & 3.261(18) & 3.24410(29) \cite{exp5}\\
	$^{209}$Bi$^{82+}$   & 5.098(7)    & 5.101(27) & 5.0840(8) \cite{exp1}\\
	\hline\hline
\end{tabular}
\end{center}
\end{table}

\end{document}